\documentclass[12pt]{amsart}
\usepackage[left=1.5in, right=1.5in, top=1.5in, bottom=1.5in]{geometry}                % See geometry.pdf to learn the layout options. There are lots.
\newtheorem{thm}{Theorem}[section]

\theoremstyle{definition}

\newtheorem{rem}[thm]{Remark}
\usepackage{graphicx}
\usepackage{amssymb}
\usepackage{amsmath}
\usepackage{epstopdf}
\usepackage{hyperref}
\title[The 1-D equilibrium shape]{The one-dimensional equilibrium shape of a crystal}
\author{Emanuel Indrei}
\address{Department of Mathematics\\
Kennesaw State University\\
Marietta, GA 30060\\
USA.}
\begin{document}
\setcounter{page}{1} 
\pagenumbering{arabic}
\maketitle

\begin{abstract}
Optimizing the free energy under a mass constraint may generate a convex crystal subject to assumptions on the potential $g(0)=0$, $g \ge 0$. The general problem classically attributed to Almgren is to infer if this is the case assuming the sub-level sets of $g$ are convex. The theorem proven in the paper is that in one dimension the answer is positive.
\end{abstract}

\section{Introduction}
The principle that the equilibrium shape of a crystal minimizes the free energy under a mass constraint was independently discovered by W. Gibbs in 1878 \cite{G} and P. Curie in 1885  \cite{Crist}. Two main elements define the free energy of a set of finite perimeter $E \subset \mathbb{R}^n$ with reduced boundary $\partial^* E$: 

$$
\mathcal{F}(E)=\int_{\partial^* E} f(\nu_E) d\mathcal{H}^{n-1}
$$
(the surface energy), where $f$ is a surface tension \footnote{A convex positively 1-homogeneous 
$f:\mathbb{R}^n\rightarrow [0,\infty)$
with $f(x)>0$ if $|x|>0$.};
and, 
$$
\mathcal{G}(E)=\int_E g(x)dx
$$
(the potential energy), where $g(0)=0$, $g \ge 0$.
The free energy is defined to be the sum
$$
\mathcal{E}(E)=\mathcal{F}(E)+\mathcal{G}(E).     
$$
\\
\noindent {\bf Problem:} If the potential $g$ is convex (or, more generally, if the sub-level sets $\{g < t\}$
are convex), are minimizers convex or, at least, connected? \cite[p. 146]{MR2807136}.\\

\noindent  A convexity assumption is in general not sufficient \cite{MR4730410}: if $n=2$ there exists $g \ge 0$ convex such that $g(0)=0$ and such that if $m>0$, then there is no solution to
\\
$$
\inf\{\mathcal{E}(E): |E|=m\}.
$$ 
\\
\noindent Nevertheless, if $n=1$ and the sub-level sets $\{g < t\}$
are convex, the convexity is true: in one dimension, the surface energy is classically the counting measure $\mathcal{H}^0$ of the set's boundary since in higher dimension the surface tension $f$ weighs the measure theoretic normal $\nu_E$ and in one dimension, for general sets, this is not well-defined, however $f(x)=|x|$ implies (when $E$ is sufficiently smooth)  
$$
\mathcal{F}(E)=\mathcal{H}^{n-1}(\partial E).
$$
Hence if $n=1$, the free energy is the sum
$$
\mathcal{E}(E)=\mathcal{H}^0(\partial E)+\int_E g(x)dx.     
$$

\begin{thm} \label{n=1}
If $n=1$,  $m \in (0, \infty)$, $g(0)=0$, $g \ge 0$, and the sub-level sets $\{g < t\}$
are convex, then 
$$
\inf\{\mathcal{E}(E): |E|=m\}
$$ 
admits minimizers and any minimizer $E_m$ is convex. 
\end{thm}

\begin{rem}
The convexity in two dimensions subject to assuming existence, local Lipschitz regularity of $g$, and some integral condition was obtained in \cite{MR4730410}. Coercivity of the potential $g$ (i.e. $g(x) \rightarrow \infty$ as $|x|\rightarrow \infty$) is sufficient for existence, nevertheless as Theorem \ref{n=1} illuminates, not in general necessary. If one considers the stronger assumption that $g$ is convex, the convexity of any minimizer for $m>0$ was shown in: (i) \cite{MR4730410} if one assumes $g$ is radial (and not identically zero; $g=0$ was investigated in \cite{MR1116536, MR1130601, MR493671}); (ii) \cite{D} with higher regularity assumptions on $f$, $g$ combined with coercivity of $g$; (iii)  \cite{MR1641031} via symmetry assumptions combined with coercivity of $g$. The three-dimensional problem is investigated in \cite{qkv}. Naturally, when minimizers are classified, one wants to study the stability problem. The sole sharp stability result in higher dimension for any mass is developed in \cite{qk} (cf. \cite{ indrei2024nonlocalalmgrenproblem} which addresses the nonlocal Almgren problem).  
\end{rem}

\begin{rem}
Assuming one investigates a strip $S \subset \mathbb{R}^2$ with width $\epsilon>0$, the one-dimensional theorem yields some information on minimizers supposing $g$ is large on $\mathbb{R}^2 \setminus S$ via taking $\epsilon \rightarrow 0^+$. Similar reasoning generates insight in $\mathbb{R}^n$.
\end{rem}

\begin{rem}
Related minimization problems are investigated in \cite{aryan2024freeenergyminimizersradial, indrei2024nonlocalalmgrenproblem, INDREI2025196, math11122670, MR4729687, I20, MR3487241}. In general, there is a way to encode a weight in the surface energy in one dimension \cite{aryan2024freeenergyminimizersradial}, however the generalization is not via a weight on a normal but on the points (i.e. historically, the tension acts on normals $\nu$ at some point $x \in \partial^* E$, $f(\nu)=f(\nu(x))$, and this is replaced via $f(x)$ where the weight is on the points). Thus in order to maintain a simple exposition, the theorem of this paper is shown with the classical counting measure.
Small mass theorems are proved in \cite{MR2807136, MR4674821, pFZ, MR4730410}.
\end{rem}

\section{The proof}

\subsection{Proof of Theorem \ref{n=1}}
Observe that 
$$
\inf\{\mathcal{E}(E): |E|=m\}
$$ 
may be constrained to sets with non-empty interior: if $E^0 =\emptyset$, then $E\subset \overline{E}=\partial E \cup E^0=\partial E$ 
and $|\partial E|\ge |E|=m>0$, which yields $\mathcal{H}^0(\partial E)=\infty$. Thus $\mathcal{E}(E)=\infty$ and $E$ is not a candidate for a minimizer. Hence let $E$ be a set having non-empty interior, and then let $x \in E$ be an interior point. If $r_a>0$ is such that $(x-r_a, x+r_a) \subset E$, note that since $0<|E|=m<\infty$, there is a point $x+r \in \partial E$, $r>0$; and, there is a point $x-l \in \partial E$, $l>0$. In particular   

\begin{equation} \label{2}
\mathcal{H}^0(\partial E)\ge 2. \\
\end{equation}
\\
\noindent {\bf Claim 1:} The sub-level sets $\{g < t\}$ are convex if and only if $g$ is non-decreasing on $[0,\infty)$ and non-increasing on $(-\infty, 0]$.\\

\noindent {\bf Proof of Claim 1:}\\

\noindent (i) If the sub-level sets $\{g < t\}$ are convex, then supposing $g$ is not non-decreasing on $[0,\infty)$, let $0<a_1<a_2$ be two elements with $g(a_1)>g(a_2)$. Set $g(a_1)>t>g(a_2)$. Then via $g(0)=0,$  $g \ge 0$, observe that $0, a_2\in \{g < t\}$, $a_1 \notin \{g < t\}$, a contradiction to the convexity of  $\{g < t\}$. Hence $g$ in non-decreasing on $[0,\infty)$. A similar argument proves $g$ is non-increasing on $(-\infty, 0]$. \\

\noindent (ii) Suppose now that $g$ is non-decreasing on $[0,\infty)$ and non-increasing on $(-\infty, 0]$. For $t>0$,  

$$
R:=\inf\{ r\ge 0: [r,\infty) \subset \{g \ge t\}\}
$$

$$
L:=\sup\{ l<0: (-\infty,l] \subset \{g \ge t\}\}
$$
\\
($\inf \emptyset =\infty$, $\sup \emptyset=-\infty$).
Thus if $-\infty<L,$ $R<\infty$, $\{g < t\}$ is one of: $(L,R)$, $[L,R)$, $(L,R]$, $[L,R]$.
Hence $\{g < t\}$ is convex.\\

\noindent Note (i) and (ii) prove Claim 1.\\
\\
Assume now that $E$ has non-empty interior and define

$$
E_+=\{x\ge0\}\cap E
$$

$$
E_-=\{x<0\}\cap E.
$$
\\
\noindent In particular, $E=E_-\cup E_+$, $|E|=|E_-|+|E_+|$.\\

\noindent {\bf Claim 2:}
$$
\int_{E_+} g(x)dx \ge \int_0^{|E_+|} g(x)dx
$$
$$
\int_{E_-} g(x)dx \ge \int_{-|E_-|}^0 g(x)dx.\\
$$
\\
\noindent {\bf Proof of Claim 2:}\\

\noindent First, assume $E_+$ is an interval. Then there exist $a_*,a$ so that $0\le a_*\le a$,
\\
$$
\int_{E_+} g(x)dx=\int_{a_*}^a g(x)dx.
$$
\\
Define 
$f(x)=g(x-a_*)$. The monotonicity of $g$ in Claim 1 thus implies 
$$
f(x)\le g(x)
$$
when $x \ge a_*$.
Hence 
$$
\int_{a_*}^a g(x)dx\ge \int_{a_*}^a f(x)dx=\int_{0}^{a-a_*} g(x)dx=\int_0^{|E_+|} g(x)dx.
$$
More generally, select sets $E_+^k=\cup_{j=1}^{n(k)} I_{j,k} \subset \{ x\ge 0\}$ such that $\{I_{j,k}\}$ is a collection of  open disjoint  intervals \footnote{Observe that in general, one may reduce to the case when $E_+=\cup_{j=1}^{n_*} Z_{j}$, where $n_*<\infty$ and $\{Z_{j}\}$ is a collection of disjoint  intervals thanks to $\mathcal{H}^0(\partial E)<\infty$. Nevertheless, since the strategy may be utilized in other problems, the approximation is more suitable.},
$$
|E_+^k \Delta E_+| \rightarrow 0.
$$
If $R>0$ and $I_R$ is the interval centered at the origin with length $2R$, 
$$
|(E_+^k \Delta E_+)\cap I_R| \rightarrow 0.
$$ 
Next, 
\begin{align*}
\int_{E_+^k \cap I_R} g(x)dx &=\int_{(\cup_{j=1}^{n(k)} I_{j,k}) \cap I_R} g(x)dx\\
&=\int_{\cup_{j=1}^{n(k)} (I_{j,k} \cap I_R)} g(x)dx=\sum_j\int_{I_{j,k} \cap I_R} g(x)dx\\
&\ge \int_{0}^{|(\cup_{j=1}^{n(k)} I_{j,k}) \cap I_R|} g(x)dx=\int_{0}^{|E_+^k\cap I_R|} g(x)dx
\end{align*}
via iterating the interval case for $I_{j,k}$. One may repeat the argument for an interval above  by possibly translating $I_{j,k}\cap I_R$ to the left: supposing $$I_{j(1),k}=(a_{j(1),k,1}, a_{j(1),k,2})$$ 
$$I_{j(2),k}=(a_{j(2),k,1}, a_{j(2),k,2}),$$  via disjointness one has without loss 
$$a_{j(1),k,2} \le a_{j(2),k,1};$$ 
in particular, there are two cases, either $a_{j(1),k,2} = a_{j(2),k,1}$ or $a_{j(1),k,2} < a_{j(2),k,1}$. When the inequality is strict, observe that one can translate $I_{j(2),k}$ to the left: define 
$f_{j,k}(x)=g(x-(a_{j(2),k,1}-a_{j(1),k,2}))$; the monotonicity of $g$ implies 
$$
f_{j,k}(x)\le g(x)
$$
when $x \in [a_{j(2),k,1}, a_{j(2),k,2}]$. Therefore
$$
\int_{a_{j(2),k,1}}^{a_{j(2),k,2}} g(x)dx\ge \int_{a_{j(2),k,1}}^{a_{j(2),k,2}} f_{j,k}(x)dx=\int_{a_{j(1),k,2}}^{a_{j(1),k,2}+(a_{j(2),k,2}-a_{j(2),k,1})} g(x)dx.
$$
In particular, one obtains the aforementioned
$$
\sum_j\int_{I_{j,k} \cap I_R} g(x)dx \ge \int_{0}^{|(\cup_{j=1}^{n(k)} I_{j,k}) \cap I_R|} g(x)dx.
$$
Now, dominated convergence implies 
$$
\int_{E_+^k \cap I_R} g(x)dx \rightarrow \int_{E_+ \cap I_R} g(x)dx
$$
$$
\int_{0}^{|E_+^k\cap I_R|} g(x)dx \rightarrow \int_{0}^{|E_+\cap I_R|} g(x)dx.
$$

\noindent Hence

\begin{equation} \label{w8}
\int_{E_+ \cap I_R} g(x)dx \ge \int_{0}^{|E_+\cap I_R|} g(x)dx.
\end{equation}
Furthermore, monotone convergence implies
\begin{equation} \label{w87}
\int_{E_+ \cap I_R} g(x)dx \rightarrow \int_{E_+} g(x)dx
\end{equation}
\begin{equation} \label{w84}
\int_{0}^{|E_+\cap I_R|} g(x)dx \rightarrow \int_{0}^{|E_+|} g(x)dx
\end{equation}
when $R \rightarrow \infty$.
Thus \eqref{w8}, \eqref{w87}, and \eqref{w84} yield
$$
\int_{E_+} g(x)dx \ge \int_{0}^{|E_+|} g(x)dx.
$$
Last, the proof of
$$
\int_{E_-} g(x)dx \ge \int_{-|E_-|}^0 g(x)dx
$$
is analogous. This proves Claim 2.\\

\noindent Claim 2 yields 
$$
\inf_{a} \int_{I+a} g(x)dx \le \int_{-|E_-|}^{|E_+|} g(x)dx \le \int_E g(x)dx
$$
where $I=(0,m)$. Since \eqref{2} implies
$$
\mathcal{H}^0(\partial E)\ge 2=\mathcal{H}^0(\partial (I+a)),
$$
note
$$
\mathcal{E}(E) \ge \inf_{a} \mathcal{E}(I+a).
$$
Next, observe that for a given potential, two cases exist: (1) $g(x)=0$ for $x \ge 0$; (2) $g(x_r)>0$ for some $x_r>0$. The first case implies that one may take any interval $(a_1,a_2) \subset \mathbb{R}_+$ such that $a_2-a_1=m$ as a minimizer; thus one may take $(0,m)$ as a minimizer. If one finds $x_r>0$ such that $g(x_r)>0$, one again has two cases: when  $g(x)=0$ for  $x\le 0$, intervals $(a_1,a_2)\subset \mathbb{R}_-$ such that $a_2-a_1=m$ are minimizers. In particular, without loss, assume $g$ is not identically zero on $\mathbb{R}_-$. Thus there is $x_l<0$ such that $g(x_l)>0$ and recall that one also has $x_r>0$ such that $g(x_r)>0$. 
If $a_k$ are numbers such that 
$$
\lim_k \mathcal{E}(I+a_k) = \inf_{a} \mathcal{E}(I+a),
$$
monotonicity yields
$$\sup_k |a_k| <\infty.$$
In order to prove this, suppose $\sup_k |a_k| =\infty$; one then may choose a subsequence $\{a_{k_i}\}$ which satisfies
$$
|a_{k_i}| \rightarrow \infty,
$$
as $i \rightarrow \infty$. Suppose that a subsequence
$$
a_{k_{i_l}} \rightarrow \infty,
$$
as $l \rightarrow \infty$. Then since $|I|=|(0,m)|=m<\infty$,  $g(x_r)>0$ for some $x_r>0$, and $g$ is non-decreasing on $\mathbb{R}_+$,
\begin{align*}
 \mathcal{E}(I)&\ge \inf_{a} \mathcal{E}(I+a) \\
 &=\lim_{l} \mathcal{E}(I+a_{k_{i_l}})\\
 &=2+\lim_l \int_{I+a_{k_{i_l}}} g(x)dx\\
 &>2+\int_{0}^{m} g(x)dx= \mathcal{E}(I),
\end{align*}
which is a contradiction. Hence there exists a subsequence 
$$
a_{k_{i_l}} \rightarrow -\infty,
$$
as $l \rightarrow \infty$. Then since there is $x_l<0$ such that $g(x_l)>0$ and $g$ is non-increasing on $\mathbb{R}_-$, a symmetric argument also yields a contradiction. In particular, there is no subsequence that satisfies
$$
|a_{k_i}| \rightarrow \infty,
$$
thus $\sup_k |a_k| =\infty$ is not true and this implies
$$\sup_k |a_k| <\infty.$$
Compactness yields a subsequence 
$$
a_{k_i} \rightarrow \alpha
$$
for an $\alpha=\alpha(m, g) \in \mathbb{R}$. 
Hence in every case one may find some $\alpha \in \mathbb{R}$ with
$$
\inf\{\mathcal{E}(E): |E|=m\}=\inf_{a} \mathcal{E}(I+a)=\mathcal{E}(I+\alpha).
$$
Now suppose $E$ is a minimizer:
$$
\inf\{\mathcal{E}(E): |E|=m\}=\mathcal{E}(E).
$$ 
Then either $E \subset \{g=0\}$ and then since $\mathcal{H}^0(\partial E) \ge 2$, $E$ is an interval. The alternative is $E \cap\{g>0\}\neq \emptyset$. Via Claim 2,
$$
\int_{-|E_-|}^{|E_+|} g(x)dx \le \int_E g(x)dx
$$

$$\mathcal{H}^0(\partial (-|E_-|, |E_+|))= 2,$$
and thus if  $E$ is not an interval, since $\mathcal{H}^0(\partial E)>2$, one may find an interval $(-|E_-|, |E_+|)$ with 
$$
\mathcal{E}((-|E_-|, |E_+|))<\mathcal{E}(E),
$$
which is a contradiction. Thus when $E_m$ is a minimizer, then there is some $\alpha$ so that one of the following is true: 
$E_m=(0,m)+\alpha$, $E_m=[0,m)+\alpha$, $E_m=[0,m]+\alpha$, $E_m=(0,m]+\alpha.$

\subsection{Optimal transport proof}
One may also prove the theorem via optimal transport theory. A key step is Claim 2 in the proof.\\
\\
\noindent {\bf Claim 2:}
$$
\int_{E_+} g(x)dx \ge \int_0^{|E_+|} g(x)dx
$$
$$
\int_{E_-} g(x)dx \ge \int_{-|E_-|}^0 g(x)dx.\\
$$
\\
\noindent {\bf Proof of Claim 2:}\\
\\
If $m_*=|E_+|>0$, $I_*=(0,|E_+|)$, consider the optimal transport $T$ which pushes $d\mu_+=\chi_{E_+\setminus I_*}dx$ forward to $d\mu_-=\chi_{I_* \setminus E_+}dx$. Observe that this can be accomplished via 
$$
|E_+ \setminus I_*|=|I_*\setminus E_+|
$$
inferred from
$$
|E_+ \setminus I_*|+|I_* \cap E_+|=|E_+|=|I_*|=|I_*\setminus E_+|+|I_* \cap E_+|.
$$
Hence 

$$
T_{\#} d\mu_+=d\mu_-
$$
thus implies
$$
\int_{E_+ \setminus I_*} g(T(x))dx=\int_{I_*\setminus E_+} g(x)dx.
$$
Next, $E_+ \setminus I_*$ is to the right of $I_*\setminus E_+$, and hence when $x \in E_+ \setminus I_*$,
$$
T(x)\le x.
$$ 
Thanks to the monotonicity of $g$,
$$
g(T(x))\le g(x).
$$
In particular,
$$
\int_{I_*\setminus E_+} g(x)dx =\int_{E_+ \setminus I_*} g(T(x))dx \le \int_{E_+ \setminus I_*} g(x)dx
$$
which then yields
\begin{align*}
\int_{I_*}g(x)dx&=\int_{E_+ \cap I_*}g(x)dx+\int_{I_*\setminus E_+} g(x)dx \\
& \le \int_{E_+ \cap I_*}g(x)dx+\int_{E_+ \setminus I_*} g(x)dx\\
&=\int_{E_+}g(x)dx.
\end{align*}
A symmetric reasoning proves
$$
\int_{E_-} g(x)dx \ge \int_{-|E_-|}^0 g(x)dx.
$$

\section{Identifying $\alpha$}
Note that the proof yields $E_m=I_a+\alpha$, with $I_a$ an interval having the form $(0,m), (0,m], [0,m), [0,m]$. 
The translation $\alpha$ in the general context depends on $m$ and $g$. If $G'(x)=g(x)$, then 
$$
\int_{I+a} g(x)dx=\int_{a}^{a+m} g(x)dx=G(a+m)-G(a).
$$
Next, the minimization 
$$
\inf_{a} \mathcal{E}(I+a)=\mathcal{E}(I_a+\alpha)
$$
immediately implies
$$
\frac{d}{da} (G(a+m)-G(a))|_{a=\alpha}=0
$$
and this yields
$$
g(\alpha+m)=g(\alpha).
$$

\noindent Examples: \\

\noindent (1) Take any increasing function $g$ on $\mathbb{R}_+$ and evenly extend it to $\mathbb{R}$. Then $
g(\alpha+m)=g(\alpha)$ readily implies $\alpha+m=-\alpha$ and one obtains $\alpha=\frac{-m}{2}$. Indeed, this may  also be inferred immediately from symmetry, however the previous computation highlights the underlying principle on identifying the translation for general potentials.\\

\noindent (2) Next, choose a continuous $g$ which (strictly) increases to $+\infty$ on $\mathbb{R}_+$ and 
(strictly) decreases from $+\infty$ to $0$ on $\mathbb{R}_-$. Let $m>0$, $\{\alpha_l, m\}=g^{-1}(g(m))$,  and define
$$
f(a):=g(m+a)-g(a),
$$
$a \in [\alpha_l,0]$. Via $\alpha_l<0$, observe through the monotonicity of $g$, 
\\
$$
f(\alpha_l)=g(m+\alpha_l)-g(\alpha_l)=g(m+\alpha_l)-g(m)<0,
$$
\\
\noindent and $g(0)=0$ yields
\\
$$
f(0)=g(m)-g(0)=g(m)>0.
$$
\\
\noindent Hence via the intermediate value theorem, there exists $\alpha \in (\alpha_l,0)$ satisfying 
\\
$$
0=f(\alpha)=g(m+\alpha)-g(\alpha).
$$
\\
\noindent Note that this yields $0\in \text{Interior}(E_m)$ for any $m>0$. To see that, first observe that since $m>0$ and $\alpha<0$, 
$$\alpha<m+\alpha<m.$$
Assuming $m+\alpha<0$ readily yields a contradiction since one may, as in the proof, translate the interval $(\alpha, m+\alpha)$ to the right to decrease the potential energy; hence $m+\alpha \ge0$ and if $m+\alpha=0$, select $\epsilon>0$ small such that 
\\
$$
\int_{-m}^{-m+\epsilon} g(x)dx =\int_\alpha^{\alpha+\epsilon} g(x)dx >\int_0^\epsilon g(x)dx
$$
\\
\noindent thanks to 
\\
$$
\frac{1}{\epsilon}\int_{0}^{\epsilon} g(x)dx \rightarrow g(0)=0
$$
$$
\frac{1}{\epsilon} \int_{-m}^{-m+\epsilon} g(x)dx \rightarrow g(-m)>0,
$$
\\
\\
when $\epsilon \rightarrow 0^+$, utilizing Lebesgue's differentiation theorem. Hence this implies that $(-m+\epsilon, \epsilon)$ generates less free energy than $(-m,0)$, which is a contradiction. In particular, $m+\alpha>0$ and this yields $0 \in (\alpha, m+\alpha)$.\\
\\

The literature has encoded information on how the minimizer employs the potential's zero level \cite{MR1641031}, however the aforementioned is rigorous under mild assumptions on $g$. If $g$ is as in the assumptions of the theorem, observe that one always finds a minimizer $E_m$ with $0 \in \overline{E_m}$ for $m>0$. Moreover, $\overline{E_m}$ is a minimizer as well whenever $E_m$ is a minimizer. But, in many cases, $0 \in \partial E_m$. Furthermore, $\alpha $ is not in general unique.

\newpage
\bibliographystyle{amsalpha}
\bibliography{References}

\end{document}